\documentclass[aps,amssymb,amsmath,pra,twocolumn,showpacs,superscriptaddress,longbibliography]{revtex4-2}

\usepackage{graphicx}% Include figure files
\usepackage{dcolumn}% Align table columns on decimal point
\usepackage{bm}% bold math
\usepackage{subfigure}
\usepackage{color}

\usepackage{amssymb,amsmath,amsfonts,latexsym,graphicx,verbatim}%,dsfont}

\usepackage[margin=0.65in]{geometry}

\usepackage[english]{babel}
\usepackage{times} 
\usepackage{dsfont}
\usepackage{latexsym}
\usepackage{fancyhdr}
\usepackage{float}
\usepackage{afterpage}
\usepackage{enumitem}
\usepackage{mleftright}

\usepackage{eso-pic, graphicx}

\usepackage{listings}
\usepackage{multirow}
\usepackage{xcolor,colortbl}

\usepackage{bbm}
\usepackage{upgreek}
\usepackage{amsmath}
\usepackage{newtxtext,newtxmath}

\definecolor{Ablue}{rgb}{0.96,0.24,0.00}

\definecolor{Abluetitle}{rgb}{0.,0.24,0.51}

\definecolor{orange}{rgb}{0.96,0.24,0.00}

\definecolor{darkred}{rgb}{0.55, 0.0, 0.0}

\definecolor{darksalmon}{rgb}{0.91, 0.59, 0.48}
\definecolor{maroon}{cmyk}{0,0.87,0.68,0.32}

\definecolor{mustard}{rgb}{1.0, 0.86, 0.35}

\definecolor{Gray}{gray}{0.85}
\definecolor{LightCyan}{rgb}{0.88,1,1}
\newcolumntype{a}{$$>$${\columncolor{Gray}}c}
\newcolumntype{b}{$$>$${\columncolor{white}}c}

\usepackage{array}
\newcolumntype{L}[1]{$$>$${\raggedright\let\newline\\\arraybackslash\hspace{0pt}}m{#1}}
\newcolumntype{C}[1]{$$>$${\centering\let\newline\\\arraybackslash\hspace{0pt}}m{#1}}
\newcolumntype{R}[1]{$$>$${\raggedleft\let\newline\\\arraybackslash\hspace{0pt}}m{#1}}

\usepackage[colorlinks=true , citecolor=black,urlcolor=blue]{hyperref}

\newcommand{\xa}{\alpha}

\newcommand{\xb}{\beta}

\newcommand{\xd}{\delta}

\newcommand{\vxe}{\varepsilon}

\newcommand{\CC}{\R{CC}}

\newcommand{\tacq}{t_{\R{acq}}}
\newcommand{\xg}{\gamma}
\newcommand{\xt}{\theta}

\newcommand{\xr}{\rho}
\newcommand{\xo}{\omega}

\newcommand{\app}{\approx}

\newcommand{\Bp}{B_{\R{pol}}}

\newcommand{\Cs}{{}^{13}\R{C}}

\newcommand{\xD}{\Delta}

\newcommand{\fr}[2]{\frac{#1}{#2}}

\newcommand{\mH}[0]{\mathcal{H}}

\newcommand{\dg}{\dagger}

\newcommand{\beq}{\begin{equation}}
\newcommand{\eeq}{\end{equation}}
                  
\newcommand{\benum}{\begin{enumerate}}
\newcommand{\eenum}{\end{enumerate}}
                    
\newcommand{\bit}{\begin{itemize}}
\newcommand{\eit}{\end{itemize}}

\newcommand{\bea}{\begin{eqnarray}}
\newcommand{\eea}{\end{eqnarray}}

\newcommand{\bev}{\begin{verbatim}}
\newcommand{\eev}{\end{verbatim}}

\newcommand{\zt}{\times}

\newcommand{\qt}{\tau}

\newcommand{\lb}{\left(}
\newcommand{\rb}{\right)}
\newcommand{\lsb}{\left[}
\newcommand{\rsb}{\right]}

\newcommand{\T}[1]{\textbf{#1}}
\newcommand{\I}[1]{\textit{#1}}
\newcommand{\R}[1]{\textrm{#1}}
\newcommand{\zl}[1]{\label{eqn:#1}}
\newcommand{\zr}[1]{Eq.\,(\ref{eqn:#1})}
\newcommand{\zfl}[1]{\protect\label{fig:#1}}
\newcommand{\zfr}[1]{\figurename\,\ref{fig:#1}}

\newcommand{\expec}[1]{\left\langle #1\right\rangle}

\newcommand{\ba}{\left\{ \begin{array}{lr}}
\newcommand{\ea}{\end{array}\right.}

\newcommand{\BRd}[1]{\textcolor{red}{#1}} %RoyalBlue MidnightBlue

\newcommand{\Tr}[1]{\textrm{Tr}\left\{{#1}\right\}}

\newcommand{\blist}[1]{
 \begin{list}{#1}%$\ast\circ\bullet\Right
 \begin{align}
	 arrow
 \end{align}
 $\checkmark\star
  { \setlength{\itemsep}{3pt}
     \setlength{\parsep}{2pt}
     \setlength{\topsep}{3pt}
     \setlength{\partopsep}{0pt}
     \setlength{\leftmargin}{1em}
     \setlength{\labelwidth}{1em}
     \setlength{\labelsep}{0.5em} } }
\newcommand{\elist}{
  \end{list}  }

\DeclareMathSymbol{\vartheta}{\mathalpha}{letters}{"12}
\DeclareMathSymbol{\theta}{\mathalpha}{letters}{"23}
\DeclareMathSymbol{\phi}{\mathalpha}{letters}{"27}
\DeclareMathSymbol{\varphi}{\mathalpha}{letters}{"1E}

\newcommand{\bef}
{
\begin{figure}[htbp]
\centering
}

\newcommand{\eef}{\end{figure}}

\addto\captionsenglish{\renewcommand{\figurename}{Fig.}}
\mleftright
\newcommand{\beginsupplement}{%
        \setcounter{table}{0}
        \renewcommand{\thetable}{S\arabic{table}}%
        \setcounter{figure}{0}
        \renewcommand{\thefigure}{S\arabic{figure}}%
				
     }

\newcommand{\affA}{Department of Chemistry, University of California, Berkeley, Berkeley, CA 94720, USA.}
\newcommand{\affB}{Chemical Science Division, Lawrence Berkeley National Laboratory, University of California, Berkeley, Berkeley, CA 94720, USA.}
\newcommand{\affC}{Fakultät Physik, Technische Universität Dortmund, D-44221 Dortmund, Germany.}
\newcommand{\affD}{Energy Geoscience Division, Lawrence Berkeley National Laboratory, Berkeley, CA 94720, USA.}
\newcommand{\affE}{Tabor Electronics Inc.  Hatasia 9, Nesher, 3660301, Israel.}

\begin{document}
%\title{\T{Minute-long Floquet prethermalization in a bulk hyperpolarized solid}}
\title{\T{Floquet prethermalization with lifetime exceeding 90s in a bulk hyperpolarized solid}}
	
	\author{William Beatrez}\affiliation{\affA}
\author{Otto Janes}\affiliation{\affA}
	\author{Amala Akkiraju}\affiliation{\affA}
	\author{Arjun Pillai}\affiliation{\affA}
	\author{Alexander Oddo}\affiliation{\affA}
	%\author{Ozgur Sahin}\affiliation{\affA}
	\author{Paul Reshetikhin}\affiliation{\affA}
	\author{Emanuel Druga}\affiliation{\affA}
	\author{Maxwell McAllister}\affiliation{\affA}
	\author{Mark Elo}\affiliation{\affE}
%	\author{$\cdots$}\affiliation{\affA}
	\author{Benjamin Gilbert}\affiliation{\affD}
\author{Dieter Suter}\affiliation{\affC}
%	\author{Jeffrey A. Reimer}\affiliation{\affF}
	\author{Ashok Ajoy}\email{ashokaj@berkeley.edu}\affiliation{\affA}\affiliation{\affB}

\begin{abstract}
We report the observation of long-lived Floquet prethermal states in a bulk solid composed of dipolar-coupled $\Cs$ nuclei in diamond at room temperature.  For precessing nuclear spins prepared in an initial transverse state, we demonstrate pulsed spin-lock Floquet control that prevents their decay over multiple-minute long periods.  We observe Floquet prethermal lifetimes $T_2'{\app}$90.9s, extended ${>}$60,000-fold over the nuclear free induction decay times.  The spins themselves are continuously interrogated for ${\sim}$10min, corresponding to the application of ${\app}$5.8M control pulses. The $\Cs$ nuclei are optically hyperpolarized by lattice Nitrogen Vacancy (NV) centers; the combination of hyperpolarization and continuous spin readout yields significant signal-to-noise in the measurements. This allows probing the Floquet thermalization dynamics with unprecedented clarity. We identify four characteristic regimes of the thermalization process, discerning short-time transient processes leading to the prethermal plateau, and long-time system heating towards infinite temperature.  This work points to new opportunities possible via Floquet control in networks of dilute, randomly distributed, low-sensitivity nuclei. In particular, the combination of minutes-long prethermal lifetimes and continuous spin interrogation opens  avenues for quantum sensors constructed from hyperpolarized Floquet prethermal nuclei. 
\end{abstract}

\maketitle

\begin{figure}[t]
  \centering
  {\includegraphics[width=0.495\textwidth]{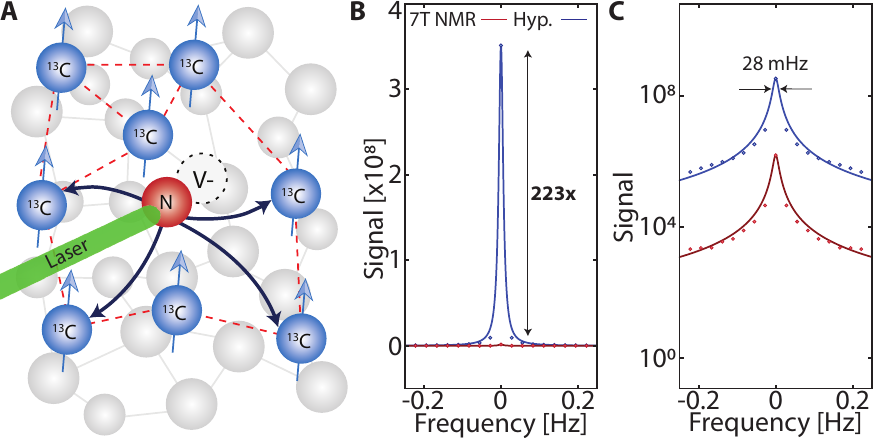}}
  \caption{\T{System.} (A) Dipolar lattice of $\Cs$ nuclei in diamond. Optically pumped NV centers are employed to hyperpolarize the $\Cs$ nuclei (blue arrows). (B-C) \I{Signal gains from hyperpolarization}, demonstrated by comparing \I{single-shot} $\Cs$ NMR spectra to conventional 7T (thermal) NMR. Data is shown in (B) linear and (C) log scales; line is a fit. Here, optical pumping was for 2min at 36mT, and thermal measurement was taken after 4hrs in the magnet. }
\zfl{fig1A}
\end{figure}

\begin{figure*}[t]
  \centering
  {\includegraphics[width=0.99\textwidth]{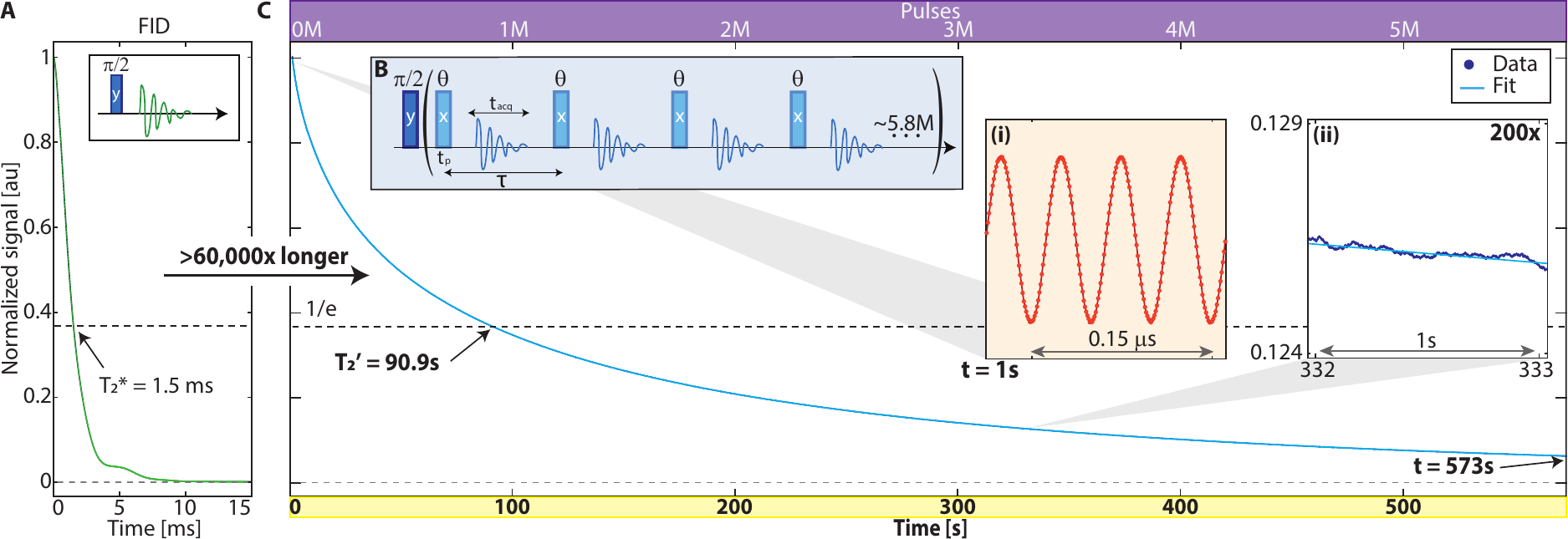}}
  \caption{\T{Floquet driving and lifetime extension.}  (A) Conventional $\Cs$ free induction decay with $T_2^{\ast}{\app} $1.5ms.  (B) \I{Floquet drive} consists of a train of $\xt$-pulses applied spin-locked with the $\Cs$ nuclei. Spins are interrogated in $t_{\R{acq}}$ windows between the pulses (blue lines), the nuclear precession is sampled every 1ns. Pulse repetition rate $\xo=\tau^{-1}$, and sequence not drawn to scale. (C) \I{Minutes-long lifetimes} of the transverse state result from the Floquet sequence ($\xt{\app}\pi/2$). Data (blue points) shows \I{single-shot} measurement of survival probability in the state $\rho_I$, and line is a fit to a sum of five exponentials. Here $t_{\R{acq}}{=}2\mu$s, $t_p{=}40\mu$s and $\qt{=}99.28\mu$s, and the 573s period corresponds to ${\app}$5.8M pulses (upper axis). We neglect here the first 100ms for clarity (see \zfr{fig2}A). \I{Inset (i)}: Raw data showing measurement of the $\Cs$ spin precession, here at 1s into the decay.  \I{Inset (ii)}: Data zoomed 200x in a 1s window. Using a 1/$e$-proxy yields $T_2’{\app}90.9$s. This corresponds to a ${>}$60,000-fold extension compared to the FID. }
\zfl{fig1B}
\end{figure*}

\T{\I{Introduction}} --  Systems pulled away from thermal equilibrium can exhibit unusual phenomena non-existent or difficult to achieve at equilibrium~\cite{Santos21}. For instance, periodically driven quantum systems can display long-lived prethermal lifetimes due to the emergence of approximately conserved quantities under the effective time-independent Hamiltonian describing the drive~\cite{Rigol14, Goldman14,Bukov15,Abanin15, Lazarides14}. For sufficiently large driving frequencies $\xo$, much higher than the intrinsic energy scales in the system Hamiltonian (hereafter $J$), these prethermal lifetimes scale exponentially with $\xo$~\cite{Kuwahara2016,Bukov15b,Abanin17, Weidinger17, Luitz20}.  Ultimately, however, the system absorbs energy and “heats up” to a featureless infinite temperature state.

The long-lived prethermal plateau and its stability against perturbations in the drive portends applications for the engineering of quantum states~\cite{Goldman14,Bukov15,Singh2019}. Fundamentally, the control afforded by periodically driven systems opens avenues to study non-equilibrium phenomena and explore novel dynamic phases of matter, some of which have no equilibrium counterparts~\cite{Else16,Khemani16}. A flurry of theoretical work has recognized Floquet prethermalization under random driving~\cite{Zhao21}, in driven linear chains~\cite{Weidinger17}, and even in the classical limit~\cite{Howell19}. Experimentally, Floquet prethermalization has been observed recently in cold-atom~\cite{Rubio20, Viebahn21, Ueda20} and NMR systems~\cite{Peng21, Yin21,Rovny18}. They demonstrated a characteristic exponential suppression of heating rates with Floquet driving. Even before the current resurgence of interest, decades-old NMR experiments had observed certain signatures of prethermalization, then referred to as “quasi-equilibrium”~\cite{Maricq87,Maricq90,Sakellariou1998,Sakellariou1999,Waugh98,Brueschweiler97}.

In this Letter, we report observation of Floquet prethermal states with lifetimes exceeding 90s at room temperature in a dipolar-coupled ensemble of $\Cs$ nuclei in diamond (see \zfr{fig1A}A). These nuclear spins, \I{randomly} positioned at 1\% concentration in the lattice, are optically hyperpolarized by interactions with NV defect centers, which enhances their polarization $\vxe{=}223$-fold with respect to the thermal limit (\zfr{fig1A}B). When placed in a Bloch transverse state $\hat{\T{x}}$ in the absence of periodic driving, these precessing nuclei naturally dephase with free induction decay lifetime $T_2^*{\app}1.5$ms and measured observables decay to zero. Under rapid pulsed spin-lock driving, however, we are able to effect a significant improvement; the observed lifetimes $T_2’{\app}$90.9s constitute a ${>}$60,000-fold extension over $T_2^*$. Moreover, with a drive consisting of ${\app}$5.8M pulses, we are able to continuously probe the thermalization process for up to 573s with high fidelity. This corresponds to ${>}10^{10}$ precession cycles of the nuclear spins. Both with respect to the number of pulses applied, and the ultimate transverse spin lifetimes, these values are amongst the largest reported in literature~\cite{Ladd2005,dong2008controlling}. Our work therefore suggests interesting opportunities for Floquet control afforded in hyperpolarizable spin networks consisting of dilute low-gyromagnetic ratio nuclei~\cite{Degen17}.

A primary contribution in this work is the ability to probe the system thermalization dynamics with unprecedented signal-to-noise (SNR). Integrated SNR (see \zfr{fig1A}B-C) exceeds $10^9$ per shot, arising from a combination of hyperpolarization and continuous spin readout in our experiments. This permits a view into the thermalization process with a high degree of clarity, in a manner not directly accessible in previous experiments. We are able to identify the four smoothly transitioning thermalization regimes that confirm theoretical predictions~\cite{Fleckenstein21b} — an initial transient to the prethermal plateau, the crossover to unconstrained thermalization and, ultimately, infinite temperature. High measurement SNR also allows characterization of heating rates over a wide range of drive frequencies. We observe system heating scaling $\propto\exp(-t^{1/2})$ at high drive frequency $\xo$. Simultaneously, the transient system response unveils interesting harmonic behavior while establishing the prethermal plateau.

\T{\I{System}} -- In a magnetic field $\T{B}_0$, the $\Cs$ nuclei interact by the dipolar Hamiltonian, $\mH_{dd} = \sum_{j<k} d_{jk}^{\CC}(3I_{jz}I_{kz} - \vec{I_j}\cdot\vec{I_k})$, with a coupling strength
$
d_{jk}^{\CC} = \fr{\mu_0}{4\pi}\hbar\xg_n^2(3\cos^2\xb_{jk}-1)\fr{1}{r_{jk}^3}
$
, where $I$ refer to spin-1/2 Pauli matrices, $\xg_n$=10.7MHz/T is the gyromagnetic ratio, and $\xb_{jk}=\cos^{-1}\lb \fr{\T{r}_{jk}\cdot\T{B}_0}{r_{jk}B_0}\rb$ is the angle of the internuclear vector $\T{r}_{jk}$ to the magnetic field. The sample is oriented with $\T{B}_0{||}[100]$, such that nearest neighbor (NN) $\Cs$ sites are decoupled. Ultimately, the median dipolar coupling is $J=\expec{d_{jk}^{\CC}}{\app}$0.66kHz (\zfr{fig1B}A). The random $\Cs$ distribution leads to a long tailed distribution in the coupling values, effectively rendering the interaction Hamiltonian \I{disordered}. In addition, the nuclei are subject to on-site disorder, i.e. local dephasing fields, $\mH_z = \sum_j c_jI_{jz}$, arising from interactions with paramagnetic impurities (e.g. P1 centers)~\cite{Reynhardt97}. At typical 20ppm P1 concentrations, $\expec{c_j^2}\app$0.4[kHz]$^2$~\cite{Ajoy19relax}. In the rotating frame of the Floquet drive, the $\Cs$ Hamiltonian is therefore $\mH = \mH_{dd} + \mH_{z}$.

Compared to previous NMR experiments, our work introduces some special features leveraging nuclear hyperpolarization~\cite{Sakellariou1998,Sakellariou1999}. The vast preponderance of NMR experiments have been limited to high-$\gamma_n$ and dense (100\% abundant) nuclei such as ${}^{19}$F,  ${}^{31}$P, and ${}^{1}$H~\cite{Rovny18, Peng21}. Instead, we focus attention to dilute networks of insensitive nuclei ($\Cs$). This provides a combination of factors critical to establishing Floquet control for long periods — \I{(i)} a relatively low $\|\mH_{\R{dd}}\|$ compared to networks constructed from sensitive (high-$\xg_n$) nuclei, scaling as $\eta^{1/2}\xg_n^2$, where $\eta$ is the nuclear enrichment, \I{(ii}) a long tailed distribution in couplings, and \I{(iii}) long nuclear $T_1$ (here ${\app}$25 min), significantly higher than many experimental systems, sets a long memory time for the nuclear states.

Indeed, these very factors, while attractive for Floquet control~\cite{Dong08}, usually make experiments challenging on account of poor sensitivity. Inductively measured nuclear signals scale $\propto \xg_n^2 $, with a measurement repetition rate set by $T_1^{-1}$, making obtaining reasonable SNR a challenge~\cite{Hoult1978}. We mitigate these difficulties by a combination of hyperpolarization and instrumental advances (allowing continuous	sampling). Hyperpolarization is carried out at $\Bp{=}36$mT through a method previously described~\cite{Ajoy17,Ajoy18}. Measurement throughput is accelerated by ${\app}\fr{1}{2}\vxe^2\lsb T_1(B_0)/T_1(\Bp)\rsb^2\frac{T_2’}{T_2^*}{\gtrsim}10^{10}$ over conventional high-field (FID-based) NMR readout.

\begin{figure}[ht!]
  \centering
  {\includegraphics[width=0.49\textwidth]{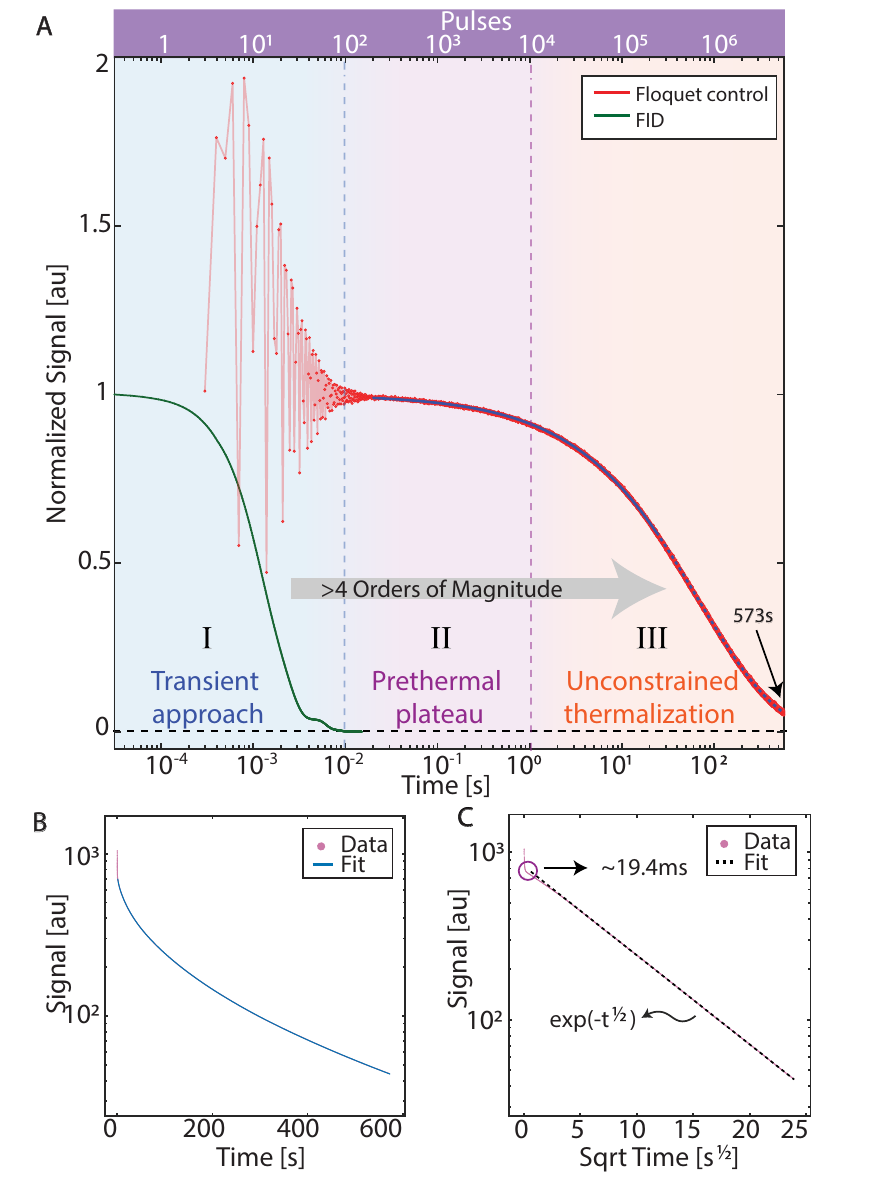}}
  \caption{\T{Floquet thermalization regimes} (A) \I{Log-scale} visualization of the full data in \zfr{fig1B}C. Points are experiment, there are ${\app}5.8$M data points here.  Lines are carried out in two segments with solid and dashed lines referring to a stretched exponential with $\xa{=}0.55$ and $\xa{=}0.5$ respectively.  Upper axis denotes number of pulses applied, here $J\tau\app$0.066.  Green points are the FID. We observe distinct, yet smoothly transitioning (shaded), thermalization regimes (\T{I}-\T{IV}): a ${\app} 10$ms oscillatory approach \T{(I)} to the Floquet prethermal plateau (\T{II}), followed by unconstrained thermalization (\T{III}). Infinite temperature regime \T{(IV)} is not reached in these measurements up to 573s. (B) \I{Semi-log} plot of the experimental data (red points) in regime \T{II}-\T{III} shows a dynamic change of thermalization rate. Moving averaging is applied here every 0.1s. Blue line is a fit to a sum of five exponentials. (C) \I{Semi-log plot against $\sqrt{t}$} yields an approximately linear dependence (dashed line) for ${\sim}500$s. Cusp (marked) at ${\app}$9.2ms marks transition to the prethermal plateau (regime \T{II}, see also \zfr{fig4}). }
\zfl{fig2}
\end{figure}

\begin{figure}[t]
  \centering
  {\includegraphics[width=0.49\textwidth]{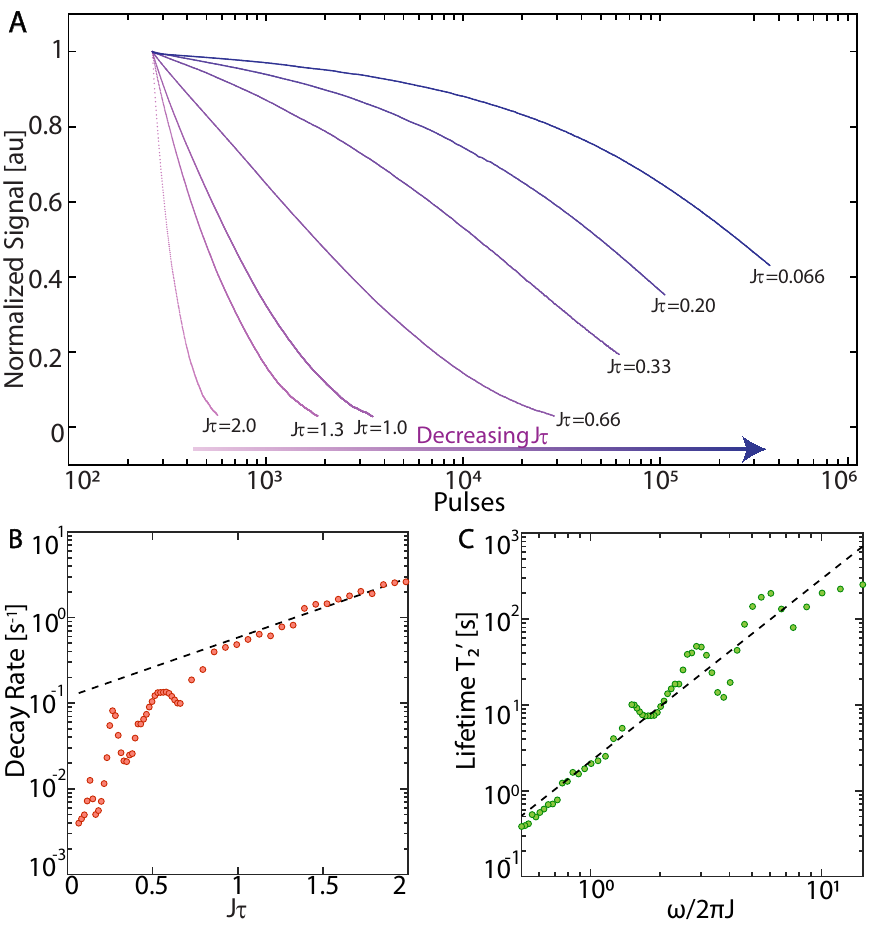}}
  \caption{\T{Exponential dependence of Floquet prethermal lifetimes.} (A) \I{Variation with $J\tau$.} Data (points) shows measured signal probing thermalization dynamics in regime \T{II}-\T{III} for representative $J\tau$ values (colorbar). Here $\xt{\app}\pi/2$ and $t_{\R{acq}}{=}32\mu$s and there are a high number (${\sim}10^3$-$10^6$) points per line~\cite{SOM}. Data is normalized at the transition points to the prethermal plateau (following \zfr{fig2}C).  The Floquet prethermal decay rates reduce considerably with decreasing $J\tau$.  See full data at Ref.~\cite{pretherm_video1}.  \I{Inset:} Zoom-in (on semilog scale).  (B) \I{Extracted decay rates} focusing on the region where decay follows ${\sim}\exp(-t^{1/2})$. Plotted in a semi-log scale against $\xo/J$, the dashed line reveals an approximately exponential scaling of the decay rates at low drive frequencies $\xo$ (dashed line is a linear fit). At high $\xo$ we observe sharp narrow features in the prethermal decay rates. (C) \I{Plotted against $J\tau$}, showing exponential scaling at higher drive frequency (dashed line). Narrow features in the decay rates superimposed on the exponential background are more emphasized here.  (D) \I{Log-scale plot} of the extracted $T_2’$ lifetimes against $\xo$. Dashed line is a linear fit. }
\zfl{fig3}
\end{figure}

\T{\I{Floquet control and measurement}} — The driving protocol is described in \zfr{fig1B}B~\cite{Ostroff66,Rhim76,Rhim78}. Post polarization, the $\Cs$ nuclei are rotated to transverse axis $\hat{\T{x}}$ on the Bloch sphere,  placing them in an initial state $\xr_I{\sim} \vxe I_x$. The Floquet drive consists of an equally spaced train of pulses of flip angle $\xt$. The center-to-center pulse separation is $\tau \lsb{=}(\xo/2\pi)^{-1}\rsb$.  After $N$ pulses, the unitary operator describing its action in the rotating frame can be written as, $U(N\tau)=\lsb\exp(i\xt I_x)\exp(i \mH \qt)\rsb^N$, where we have made a simplifying assumption of $\xd$-pulses. The data is sampled after every pulse, $t_j{=}j\qt$, and the evolution can be described by the operation $U(t)=\prod_{j=1}^{N}\exp(i\mH^{(j)}\qt)$, where we refer to the toggling frame Hamiltonians after every pulse~\cite{Haeberlen76}, $\mH^{(j)} = \exp(ij\xt I_x) \mH  \exp(-ij\xt I_x)$. This evolution can be recast as, $U(t)=\exp(i\mH_F N\qt)$, where $H_F$ is the Floquet Hamiltonian that captures the system dynamics under the drive.  $\mH_F$ can be expanded in a Floquet-Magnus expansion~\cite{Magnus54,Wilcox67,Blanes09} to leading order in the parameter $\zeta{=}2\pi J/\xo$, and in the regime $\zeta{\ll}1$, yields a time independent Hamiltonian,
\beq
{\mH}^{(0)}_{F}= \sum_{j=1}^{N}\mH^{(j)} \app \sum_{j<k}d_{jk}^{\CC}\lb \fr{3}{2}\mH_{\R{ff}}-\vec{I_j}\cdot\vec{I_k}\rb,
\zl{H0}
\eeq 
with the flip-flop Hamiltonian, $\mH_{\R{ff}} = I_{jz}I_{kz}+ I_{jy}I_{ky}$~\cite{Ajoy20DD}. The $\mH_z$ dephasing fields are filtered out in ${\mH}^{(0)}_{F}$. For sufficiently small $\zeta$,  \zr{H0}  holds irrespective of the flip-angle $\xt$, except for certain special values ($\xt\app \pi, 2\pi$). We note that this constitutes a key difference with respect to conventional dynamical decoupling control (CPMG~\cite{Carr54}), wherein the interspin couplings are retained and result in rapid $\Cs$ decay~\cite{Ajoy20DD}.  The higher order terms in the Magnus expansion are progressively smaller, but contribute to long time system dynamics~\cite{Wilcox67,Ernst}. Importantly, the initial transverse magnetized state $\xr_I$ is a conserved quantity under ${\mH}^{(0)}_{F}$, since $[\xr_I{,}{\mH}^{(0)}_{F}]{=}0$. This leads to prethermal lifetimes that depend exponentially on the drive frequency $\xo$. Ultimately, the divergence of the expansion manifests in the system heating to infinite temperature.

\zfr{fig1B}C shows the measured survival probability $F(N\qt)$ of the state $\xr_I$ under the applied Floquet drive.  This can be expressed as, $F(N\qt)=\fr{1}{2}\Tr{\xr_IU(N\qt)^{\dg}\xr_IU(N\qt)}$. We have neglected the first 100ms here for clarity (see \zfr{fig2}A for full data). Data shows significant extension in the transverse state lifetimes. Points in \zfr{fig1B}C are the experimental data while the line is a fit to a sum of five exponentials (zoomed in \zfr{fig1B}C\I{(ii)}); the high measurement SNR is evident in the zoomed data. The product $J\tau$ is a convenient metric to label the Floquet regime of operation, and in these measurements $J\tau$=0.066. The $\xt{\app}\pi/2$ pulses here are applied every $\qt{\app}100\mu$s, and the 573s period encapsulates ${\app}$5.8M pulses. For comparison, the conventional $\Cs$ free induction decay~\cite{Lowe1959} in the absence of Floquet driving is shown in \zfr{fig1B}A, where decay occurs in $T_2^{\ast}{\app}$1.5ms on account of internuclear couplings and static field disorder. High SNR and continuous weak measurement readout allows us to recognize (see \zfr{fig2}B) a dynamic change in the decay rate constant along the curve, making it difficult to quantify the decay rate by a single number. The data especially past 100ms is found to fit well to the stretched exponential ${\sim}\exp\lsb -(t/T_2’)^{1/2}\rsb$, from where we extract $T_2’{\app}$353s. Alternatively, using a 1/$e$- intersection (dashed line in \zfr{fig1B}C) as a convenient proxy yields, $T_2’{=}90.9$s.  The extension leads to substantial line-narrowing of the $\Cs$ NMR spectrum (${\sim}28$mHz in \zfr{fig1A}C).

The measurement procedure for \zfr{fig1B}C is detailed in the Supplementary Information~\cite{SOM}. The signal is sampled every 1ns in $\tacq$ windows between the pulses (see \zfr{fig1B}B). Such continuous readout (akin to weak measurement~\cite{Pfender19}) yields significant SNR advantages over point-by-point stroboscopic measurements. Rapid data sampling throughput (at $f_s{=}\qt^{-1}$) also allows further filtering to be applied when the dynamics are slow compared to $f_s$. With this, we obtain a single-shot SNR ${>}10^3$ per measurement point, and ${\app} 4\zt 10^9$ for the integrated signal (see \zfr{fig1A}C).

\begin{figure}[t]
  \centering
  {\includegraphics[width=0.495\textwidth]{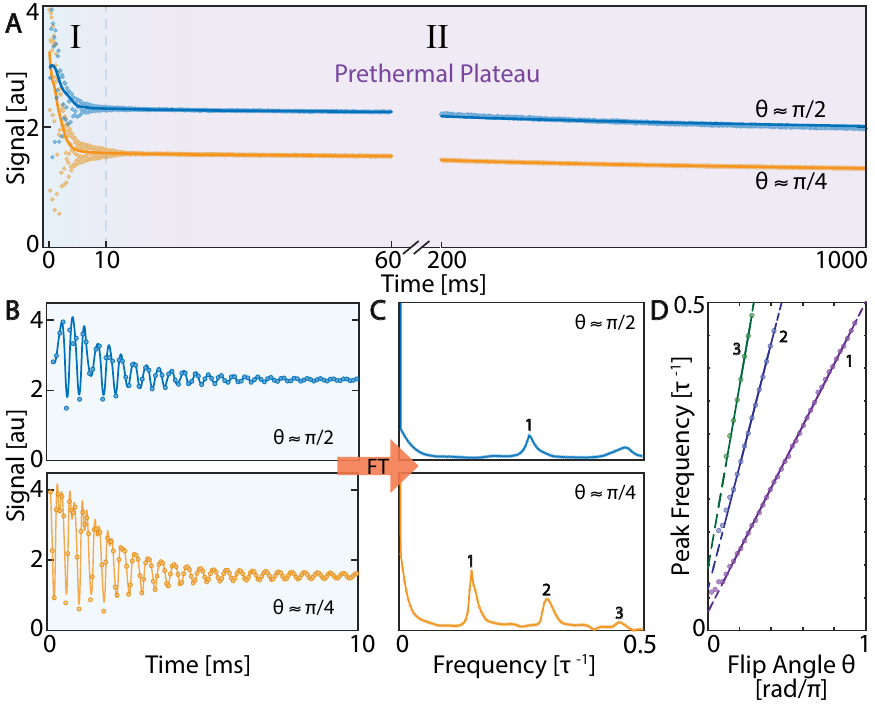}}
  \caption{\T{Transient approach to prethermal plateau.} (A) \I{Oscillations in the approach to prethermal plateau} seen for data zooming in on region \T{I}, in a 1s long window (see \zfr{fig2}A). Data (points) corresponds to $\xt{\app}\{\pi/2,\pi/4\}$ respectively.  Solid line is data with moving average filtering applied over the entire region (see \zfr{fig2}B).  (B) \I{Zoom in} to region \T{I} shows the transient approach with high SNR. It is evident that the oscillations are at higher frequency for $\xt{\app}\pi/2$.  Solid line is a spline fit to guide the eye.  (C) \I{Fourier transforms} of panels in B allows identification of the frequency components constituting the oscillations as a function of $\xo{=}\tau^{-1}$. Harmonics are represented by numbers. It is clear that primary oscillation frequency is higher for $\xt{\app}\pi/2$, where we extract the primary harmonic position at ${\app}0.26\qt^{-1}$. (D) \I{Variation with flip angle} $\xt$. Data shows the position of the oscillation frequency for the primary and higher harmonics (numbered). See full data at Ref.~\cite{pretherm_video2}. Solid lines are linear fits, while dashed line is an extrapolation. Slopes are in the ratio expected. }%Deviation at $\xt{\rt}0$ arises from non-linear scaling of the flip angles at short pulse lengths due to pulse transients. }
\zfl{fig4}
\end{figure}

\T{\I{Floquet Prethermalization}} -- To better illustrate thermalization dynamics of the spins, \zfr{fig2}A shows the full data on a logarithmic time scale. The FID is also shown, and lifetime extension is evident from the shift in the curves. Points are experimental data with no moving average applied, and the solid and dashed lines are stretched exponential fits. We identify distinct, albeit smoothly transitioning, regimes in the thermalization process (shaded in \zfr{fig2}A). Following Ref.~\cite{Rigol14}, we refer to them as: \T{(I)} an initial regime of constrained thermalization ($0{<}t{<}20$ms), where we observe oscillatory behavior with a harmonic frequency response of the Floquet drive frequency $\xo$, (\T{II}) the prethermal plateau, leading into (\T{III}) unconstrained thermalization towards the (\T{IV}) infinite temperature state (not reached in these experiments).

Let us first focus our attention to the dynamics in regimes \T{II} and \T{III}. \zfr{fig2}B-C shows two complementary visualizations after moving average filter is applied over the entire data. \zfr{fig2}B, plotted on a semi-log scale, makes evident that the decay rate constant changes over the entire thermalization period. The high SNR and rapid sampling rate, however, allows us to unravel the exact rate change behavior in a manner not accessible in previous experiments. It is easiest seen when re-plotted against $\sqrt{t}$ in \zfr{fig2}C, where we obtain an approximately linear trend (dashed line) over a long period (${\sim}$500s). The prethermal dynamics is therefore ${\sim}\exp(-t^{\alpha})$ with exponent $\alpha{\app}1/2$. Decades-old NMR experiments had observed a similar trend in paramagnetic impurity rich solids~\cite{Tse68,Lin73}. We emphasize however the high SNR of the data in \zfr{fig2}, proffering insights into, and deviations from, this behavior. At higher $J\tau$ values, for instance, we observe a dynamic decrease in $\alpha$ away from 1/2 in regime \T{III} (movie available at Ref.~\cite{pretherm_video1}). The turning point (cusp) in data in \zfr{fig2}C, obtained after moving average filtering over the oscillations in regime \T{I}, also allows a convenient means to quantify the exact point of transition to prethermal plateau. The length of this period ($\app$10-20ms) closely mirrors the period over which the FID completely decays (see \zfr{fig2}A).

To study the scaling of the prethermal lifetimes with the frequency of the Floquet drive $\xo$, \zfr{fig3}A shows similar data at a range of $J\tau$ values. This is carried out by varying the inter-pulse spacing $\tau$ in \zfr{fig1B}B. The full dataset (shown in Supplementary Information) consists of measurements at 57 such $J\tau$ values, but we show a restricted set here for clarity. Again, there is a high density of data points in each experimental line. To restrict attention to regions \T{II}-\T{III}, we normalize the data at the transition points to the prethermal plateau, identified from the cusps as in \zfr{fig2}C. The data show thermalization proceeding more slowly for lower values of $J\tau$. The dynamic change of rate coefficient makes plotting a single graph that encapsulates the full long-time behavior difficult. Instead, we extract the decay rates focusing on regime \T{II}, where decay (similar to \zfr{fig2}C) follows an exponent  $\alpha{\app}1/2$.

This is presented in two complementary viewpoints in \zfr{fig3}B-C. First, in \zfr{fig3}B plotted on a semi-log scale with respect to the drive period $\tau$, we see a linear trend in the decay rates, especially at high $\tau$ (dashed line). This points to an approximately exponential scaling of the state preservation lifetimes with drive frequency, one of the signatures of Floquet prethermalization. At low $\tau$ however, we observe a flatter slope with sharp features in the decay rates. \zfr{fig3}C shows an alternate view instead in terms of $\xo$. Extracting the transverse state lifetimes $T_2’$ on the log-log plot, we find a slope of $2.1\pm0.1$ at low-frequency suggesting a Fermi's Golden Rule scaling with drive frequency~\cite{Fleckenstein21a}.

The sharp peaks in the decay rates in the high $\xo$ regime in \zfr{fig3}B-C are intriguing. We believe this is a manifestation of quantum sensing — the $\Cs$ nuclei see an enhanced decay rate when subjected to environmental magnetic fields at a fixed frequency $f_\R{ac}$ matched in periodicity (resonant) with the pulse sequence,  at $f_\R{ac}{=}\xt/(2\pi\qt)$. The first two peaks are observed at $f_\R{ac}{\app}2.5$kHz and $f_\R{ac}{\app}5.0$kHz. This is possible because the pulsed spin-lock sequence exhibits dynamical decoupling properties similar to quantum sensing protocols~\cite{Degen17}. The exact origin of these fields in \zfr{fig3}B-C are unclear and beyond the scope of the current manuscript. A more detailed exposition on exploiting Floquet prethermal states for quantum sensing will be presented elsewhere.

\T{\I{Approach to prethermal plateau}} -- Finally, let us elucidate how the nuclear spins approach the Floquet prethermal plateau~\cite{Haldar18}, focusing attention on regime \T{I} of \zfr{fig2}A. We observe transients in the survival probability leading into the plateau; this is shown for two choices of the flip-angle $\xt$ in \zfr{fig4}A ($\xt{\app}\pi/2$ and $\xt{\app}\pi/4$) respectively. High SNR allows us to track the oscillatory dynamics after every pulse, providing a window into how the approximately time-independent Hamiltonian is established. Moreover, the prethermal plateau level is itself dependent on $\xt$.

The transients last for $t{\app}10$ms, which is approximately the total lifetime for the original FID, and is of the order of magnitude of $\|\mH_{dd}\|^{-1}$ (see \zfr{fig2}A). As \zfr{fig4}B indicates, the oscillation periodicity is closely related to the flip angle employed; for $\xt{\app}\pi/2$, for instance, the oscillations occur at a fourth of the frequency of the Floquet drive $\xo$. To see this more clearly, \zfr{fig4}C shows the respective Fourier transforms in a 10ms region. Plotted against $\xo$, we identify harmonics of the oscillatory dynamics (numbers). For $\xt{\app}\pi/4$ (lower panels in \zfr{fig4}B-C), we recognize a primary harmonic and higher harmonics at ${\app} n\xo/8$, where $n$ is an integer.

Intuitively, this characteristic periodicity can be thought of as arising from the number of pulses $N_k$ required to return the Floquet unitary to a prior configuration; \I{i.e.} such that the toggling frame Hamiltonian after $2N_k$ pulses is equivalent to that after $N_k$,  $\mH^{(2N_k)} {=} \mH^{(N_k)}$.  This corresponds to effectively completing a $2\pi$ rotation of the Hamiltonian in the toggling frame.  Four pulses are therefore needed for $\xt{=}\pi/2$ in \zfr{fig4}A. In general, the primary harmonic frequency is expected to be at frequency $f{=}\xt/(2\pi\tau)$. Experiments confirm this picture; we extract the oscillation frequencies in regime \T{I} as a function of $\xt$, and they fall neatly onto three straight lines for the three harmonics (see \zfr{fig4}D). We hypothesize that the higher harmonics arise from bilinear and trilinear terms in the density matrix produced by dipolar evolution.  The experimentally measured slopes are in the ratio $1{:}1.98{:}2.93$, close to the $1{:}2{:}3$ pattern expected.

In conclusion, we have observed Floquet prethermalization of dipolar-coupled nuclear spins in a bulk solid at room temperature. The observed ${>}$90s-long prethermal lifetimes in diamond $\Cs$ nuclei are over four orders of magnitude longer than free induction decay times, and significantly longer than in other systems. Our measurements unveil regimes of thermalization with a degree of clarity not accessible in previous NMR studies. Apart from fundamental insights, our work points to attractive opportunities possible via Floquet control in hyperpolarizable, dilute and low-$\gamma_n$ nuclear networks. Protection and continuous interrogation of spins along a Bloch transverse axis for ${\sim}$10min periods opens avenues for high-sensitivity magnetometers, gyroscopes~\cite{Ajoy12g,Ledbetter12}, and spin sensors~\cite{Abobeih19} constructed out of hyperpolarized prethermal $\Cs$ nuclei.

We gratefully acknowledge M. Markham (Element6) for the diamond sample used in this work, and discussions with S. Bhave, M. Bukov, C. Fleckenstein, C. Meriles, J. Reimer, D. Sakellariou, and A. Souza. This work was funded by ONR under contract N00014-20-1-2806. BG was supported by DOE BES CSGB under contract DE-AC02-05CH11231.

% \bibliography{./Biblio}

\pagebreak

\clearpage
\onecolumngrid
%\begin{widetext}
\begin{center}
\textbf{\large{\textit{Supplementary Information} \\\smallskip
Floquet prethermalization with lifetime exceeding 90s in a bulk hyperpolarized solid}}\\
\hfill \break
\smallskip
William Beatrez,$^{1}$  Otto Janes,$^{1}$ Amala Akkiraju,$^{1}$ Arjun Pillai,$^{1}$ Alexander Oddo,$^{1}$ Paul Reshetikhin,$^{1}$ \\ Emanuel Druga,$^{1}$ Maxwell McAllister,$^{1}$ Mark Elo,$^{2}$ Benjamin Gilbert,$^{3}$ Dieter Suter,$^{4}$ and Ashok Ajoy,$^{1,5,\BRd{\ast}}$\\
\smallskip
\emph{${}^{1}$ {\small Department of Chemistry, University of California, Berkeley, Berkeley, CA 94720, USA.}}
\emph{${}^{2}$ {\small Tabor Electronics Inc.  Hatasia 9, Nesher, 3660301, Israel.}}
\emph{${}^{3}$ {\small Energy Geoscience Division, Lawrence Berkeley National Laboratory, Berkeley, CA 94720, USA.}}
\emph{${}^{4}$ {\small Fakultät Physik, Technische Universität Dortmund, D-44221 Dortmund, Germany.}}
\emph{${}^{5}$ {\small Chemical Science Division, Lawrence Berkeley National Laboratory, University of California, Berkeley, Berkeley, CA 94720, USA.}}
\end{center}

\twocolumngrid

\beginsupplement

\begin{figure}[t]
  \centering
  {\includegraphics[width=0.49\textwidth]{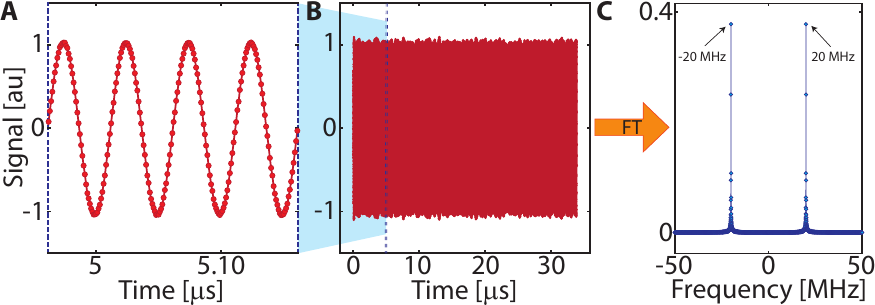}}
  \caption{\T{Data processing chain. }(A) Raw data of $\Cs$ nuclear precession measured inductively, here heterodyned from 75MHz (Larmor frequency) to 20MHz. The window shown is a 150ns part of the larger acquisition window. (B) Raw data corresponding to one complete acquisition period between the pulses, here $\tacq{=}32\mu$s. (C) Fourier transformation reveals characteristic peaks at the heterodyning frequency ${\pm}20$MHz. This amplitude forms the primary data that is plotted in the main paper. We therefore obtain one data point per acquisition window, and in \zfr{fig1B}C a total of ${\app}$5.8M data points. }
\zfl{Sfig1}
\end{figure}

\section{Data Processing}
We now present  details of the data processing employed in this manuscript. \zfr{Sfig1} describes the chain of steps involved in obtaining decay curves such as \zfr{fig1B}C of the main paper. We readout the NMR signal continuously in $t_{\R{acq}}$ periods between the pulses, and the pulses are spaced apart by $\qt{=}100\mu$s. A representative such acquisition window is shown in \zfr{Sfig1}B, in this case $t_{\R{acq}}{=}32\mu$s. The data here (taken 1s into the decay) is sampled at 1Gs/s ($\xD t$=1ns). A zoom in (\zfr{Sfig1}A) reveals high SNR oscillations corresponding to the precession of the hyperpolarized nuclei; the frequency here is 20MHz (heterodyned from the 75MHz Larmor frequency). For each such window, we take a Fourier transform (\zfr{Sfig1}C), and extract the 20MHz peak. This corresponds effectively to digital bandpass filtering with a filter linewidth of ${\sim}t_{\R{acq}}^{-1}$. Each such point is then plotted to create the decay curves in \zfr{fig1B}C, \zfr{fig2} and \zfr{fig3}. Since the measurements are carried out after every pulse, we obtain a data point in \zfr{Sfig1}D every $\qt{=}100\mu$s. Over the 573s decay period, this corresponds to ${\app}5.8$ million measurement points. However, the data itself is slowly varying (except in regime \T{I}), and can be thought of as being effectively oversampled by the measurement points. Moving average filtering thus increases SNR further, acting as a low-pass filter to suppress higher frequency variations. We typically employ a moving average filter size of 0.1s.

\begin{figure}[t]
  \centering
  {\includegraphics[width=0.4\textwidth]{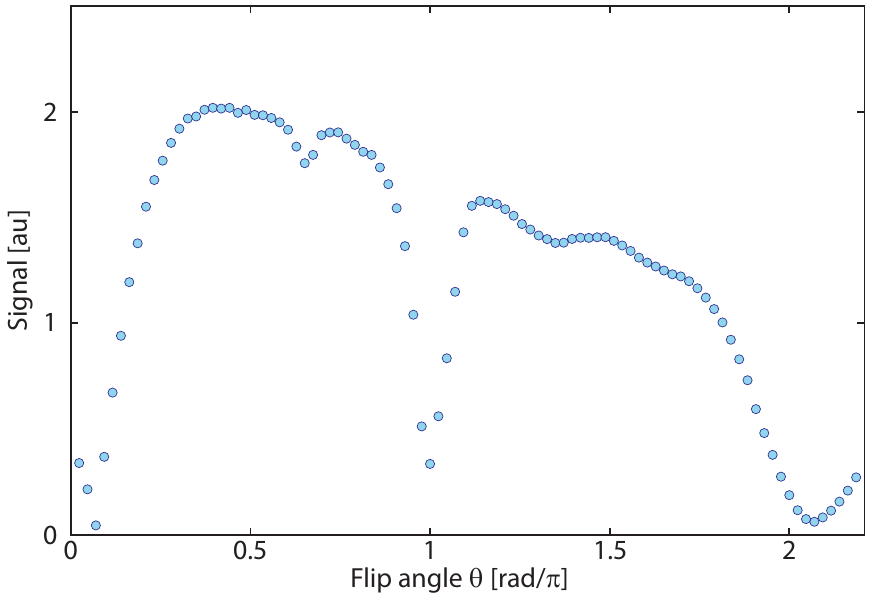}}
  \caption{\T{Signal dependence on flip angle $\xt$}. Panel shows the net signal obtained upon application of a train of ${\xt}$-pulses following the initial $\pi/2$ pulse. We identify clear signal dips at $\xt{\app} \{\pi,2\pi\}$ corresponding to rapid decay due to evolution under the dipolar Hamiltonian (see \cite{Ajoy20DD}). In these experiments $\tacq{=}32\mu$s and pulse spacing $\qt{=}100\mu$s. }
\zfl{Sfig2}
\end{figure}

\begin{figure}[t]
  \centering
  {\includegraphics[width=0.49\textwidth]{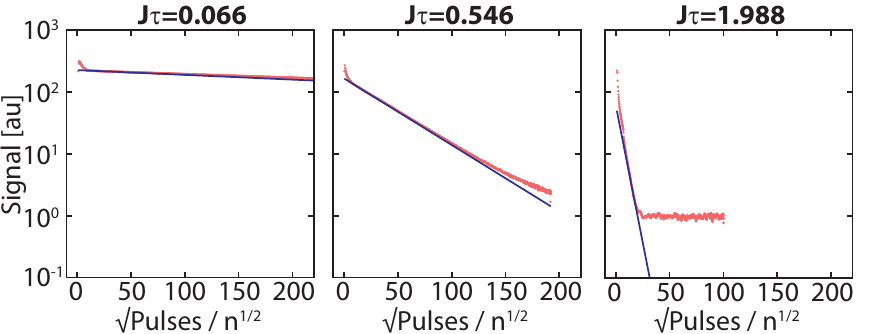}}
  \caption{\T{Raw data corresponding to \zfr{fig3}} of the main paper. Movie (see at YouTube here:\cite{pretherm_video1}) shows signal plotted on a semi-log scale against$ \sqrt{t}$. Data are red points, while the blue line is a fit taken in the region where $\xa{\app} 1/2$. The extracted rates are plotted as points in the lower panels of \zfr{fig3} of the main paper.}
\zfl{Sfig3}
\end{figure}

\begin{figure*}[t]
  \centering
  {\includegraphics[width=0.8\textwidth]{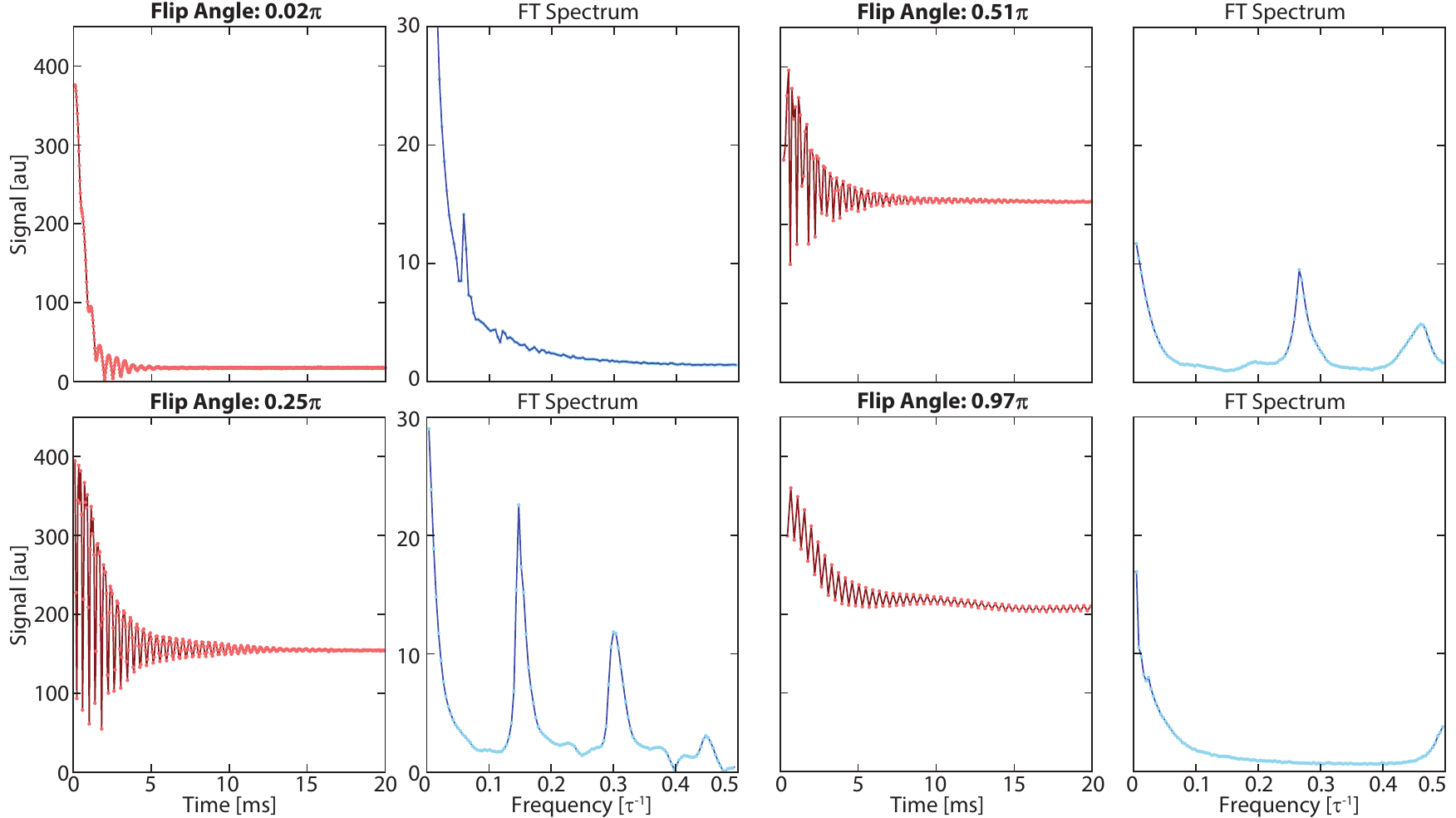}}
  \caption{\T{Raw data corresponding to \zfr{fig4}} of the main paper. Movie (see at YouTube here:\cite{pretherm_video2}) shows frames corresponding to the transient approach to prethermal plateau for different flip angles $\xt$. Panels show the transient approach and Fourier transform of the oscillations showing characteristic frequencies. }
\zfl{Sfig4}
\end{figure*}

\section{Materials and Methods}
The sample used in these experiments consists of a CVD fabricated single crystal of diamond with ${\sim}$1ppm of NV centers. The sample is placed flat, i.e. with its [100] face pointing parallel to the hyperpolarization and interrogation magnetic fields (36mT and 7T respectively). In this configuration, the internuclear vector between $\Cs$ nuclei at NN sites on the lattice are positioned at the magic angle, and hence are  suppressed. 

For hyperpolarization, we employ continuous optical pumping and swept microwave irradiation for ${\sim}$40s through a technique described previously. Hyperpolarization is carried out at low field,  the sample is shuttled to high field, and the Floquet sequence in \zfr{fig1B}B is then applied. NV-$\Cs$ polarization transfer occurs via biased Landau-Zener traversals in the rotating frame; spin diffusion serves to transfer polarization to bulk $\Cs$ nuclei in the diamond lattice. 

\zfr{Sfig2} shows the variation of the integrated signal as a function of the flip angle $\xt$ employed in the pulse sequence.  We refer the reader to Ref. \cite{Ajoy20DD} for a more detailed exposition of the observed trend. For experiments \zfr{fig1B} and \zfr{fig2} of this paper, we employ a pulse duty cycle of $\sim$50\%, where the measured SNR is highest.

The figure movie \zfr{Sfig3} (accessible in Ref. \cite{pretherm_video1}) shows the full dataset corresponding to \zfr{fig3} of the main paper. Similar to \zfr{fig3}B, we plot the data against $\sqrt{t}$ in a semi-log axis. Here the value $J\tau$ is varied by altering the spacing between the pulses in \zfr{fig3}B. The blue straight lines in the movie show the fit to $\alpha{=}$1/2 region.  These decay rates, corresponding to the slope of the blue lines in Ref. \cite{pretherm_video1}, are plotted in the lower panels of \zfr{fig3}.

Similarly, the figure movie \zfr{Sfig4} (accessible in Ref. \cite{pretherm_video2}) shows the full dataset corresponding to \zfr{fig4} of the main paper, describing the approach to the Floquet prethermal plateau for different values of the flip angle $\xt$.

\end{document}